\documentclass[11pt]{article}

\usepackage{amsfonts,amsthm}
\usepackage{amsmath}  
\usepackage{latexsym}
\usepackage{amssymb}
\usepackage{mathrsfs}
\usepackage{tikz}

\newtheorem{defi}{Definition}[section]

\newtheorem{prop}{Proposition}[section]

\newtheorem{rem}{Remark}

\newcommand{\lda}{\lambda}

\newcommand{\ZZ}{\mathbb{Z}}

\newcommand{\CC}{\mathbb{C}}

\newcommand{\1}{\mbox{\hspace{.0em}1\hspace{-.24em}I}}

\begin{document}

\begin{center}
\textbf{\Large Classical $N$-Reflection Equation and Gaudin Models}\\[3ex]
\large{Vincent Caudrelier$^a$, Nicolas Cramp\'e$^b$}\\[3ex]

$^a$ School of Mathematics, University of Leeds, LS2 9JT, UK  \\[3ex]

$^b$ Laboratoire Charles Coulomb (L2C), Univ Montpellier, CNRS, Montpellier, France.\\[3ex]
 
\end{center}

\vspace{0.5cm}

\centerline{\bf Abstract}  
We introduce the notion of $N$-reflection equation which provides a generalization of the usual classical reflection equation describing 
integrable boundary conditions. The latter is recovered as a special example of the $N=2$ case. The basic theory is established and illustrated 
with several examples of solutions of the $N$-reflection equation associated to the rational and trigonometric $r$-matrices. A central result 
is the construction of a Poisson algebra associated to a non skew-symmetric $r$-matrix whose form is specified by a solution of the $N$-reflection 
equation. Generating functions of quantities in involution can be identified within this Poisson algebra. As an application,
we construct new classical Gaudin-type Hamiltonians, particular cases of which are Gaudin Hamiltonians of $BC_L$-type.

\section{Introduction}

Classical integrable systems have been formulated in terms of the classical r-matrix in \cite{Skly,STS1,FT} for which the central equation is
called the classical Yang-Baxter equation
\begin{eqnarray}\label{eq:cybe}
[\ r_{ab}(\lda,\mu)\ ,\ r_{ac}(\lda,\nu)\ ]+ [\ r_{ab}(\lda,\mu)\ ,\ r_{bc}(\mu,\nu)\ ]-[\ r_{ac}(\lda,\nu)\ ,\ r_{cb}(\nu,\mu)\ ]=0\,.
\end{eqnarray}
Then, the study of boundary conditions which preserve the integrability of the model leads to the classical reflection equation \cite{Sk} 
\begin{eqnarray}
 r_{ab}(\lda,\nu) \  k_a(\lda)k_b(\nu) +  k_a(\lda)\  r_{ba}(\nu,-\lda) \ k_b(\nu) -  k_b(\nu)\  r_{ab}(\lda,-\nu)\ k_a(\lda)    -   k_a(\lda) k_b(\nu)\  r_{ba}(-\nu,-\lda)=0\;.\label{eq:RE}
\end{eqnarray}
This equation appears naturally in classical integrable systems based on the $BC_L$ root system and can be interpreted via a $\ZZ_2$ action on the $A_{2L}$ root system. This point of view on integrable boundary conditions, 
sometimes called ``folding'' for short, has been used extensively e.g. in \cite{CC2,CZ1,CZ2}.
The search for integrable systems associated to the action of more complicated finite groups, and in particular of the cyclic group $\ZZ_N$, has attracted a lot of attention recently 
(see e.g. \cite{poly,DO,CY,CY2,CC,EF,VY,CS,CF}).

In this paper, we introduce a new equation which generalizes the classical reflection equation \eqref{eq:RE} as well as algebraic structures related to $\ZZ_N$ (cyclotomic) models. 
In Section \ref{N_ref}, we define the $N$-reflection equation and prove that it allows one to construct new solutions of the 
classical Yang-Baxter equation from old ones. Then, in Section \ref{sec:solu}, we show on some examples that 
this equation has interesting solutions and, in particular, solutions where the action on the spectral parameters is a Mobi\"us transformation. In Section \ref{sec:alg}, we show how the $N$-reflection equation allows one 
to define a certain Poisson subalgebra of a linear Poisson algebra defined by a classical $r$-matrix and use this to obtain new integrable Gaudin models. We also show that the Hamiltonian equations of motion generated by the elements in involution in the Poisson subalgebra can be written in Lax form and give an explicit formula for the second matrix of the Lax pair.

\section{Classical $N$-reflection equation}\label{N_ref}

In this paper, we consider a solution $r_{ab}(\lda,\mu)$ of the classical Yang-Baxter equation \eqref{eq:cybe} acting on $\CC^n \otimes \CC^n \otimes \CC^n$. We now introduce the following generalization of the classical reflection equation associated to $r$.
\begin{defi}
	Let $\tau$ and $g^{(j)}, j=0,\dots,N-1$ be functions of the spectral parameter,  $\tau^j$ be defined recursively by $\tau^j(\nu)=\tau(\tau^{j-1}(\nu))$ with $\tau^0(\nu)=\nu$ and
	\begin{equation}
	k^{(j)}(\nu)=k^{(j-1)}(\nu) k(\tau^{j-1}(\nu))\quad\text{with}\quad k^{(0)}=\1_n\;.
	\end{equation}
		The classical $N$-reflection equation for $\tau$, $g^{(j)}$ and the matrix $k$ is defined by
		\begin{eqnarray}
		&&\sum_{j=0}^{N-1}\  g^{(j)}(\nu)\  k^{(j)}_b(\nu)\  r_{ab}(\lda,\tau^j(\nu)) \ \left( k^{(j)}_b(\nu)\right)^{-1} \ \  k_a(\lda) \nonumber\\
		&&= k_a(\lda) \  \sum_{j=0}^{N-1}\ g^{(j)}(\nu)\  k^{(j)}_b(\nu)\  r_{ab}(\tau(\lda),\tau^j(\nu)) \ \left( k^{(j)}_b(\nu)\right)^{-1}\;.\label{eq:NRE}
		\end{eqnarray}
Without loss of generality, we can set $g^{(0)}(\nu)=1$. 
\end{defi}
For the sake of comparison with the usual reflection equation, let us write explicitly the classical 2-reflection equation by multiplying by $k_b(\nu)$ on the right hand side:
\begin{eqnarray}
 && r_{ab}(\lda,\nu) \  k_a(\lda)k_b(\nu)  + g^{(1)}(\nu)\  k_b(\nu)\  r_{ab}(\lda,\tau(\nu))\ k_a(\lda) \nonumber\\
 &&\hspace{2cm} =  k_a(\lda)\  r_{ab}(\tau(\lda),\nu) \ k_b(\nu)  + g^{(1)}(\nu)\   k_a(\lda) k_b(\nu)\  r_{ab}(\tau(\lda),\tau(\nu))\;.\label{eq:NRE2}
\end{eqnarray}
Then, we have the following results establishing the connection with the usual reflection equation. 
\begin{prop}
Suppose the $r$-matrix depends only on the difference of the spectral parameters and is skew-symmetric, taking $g^{(1)}(\nu)=-1$ and $\tau(\nu)=-\nu$, the classical 2-reflection equation \eqref{eq:NRE2} becomes the usual classical reflection equation
\begin{eqnarray}
\label{usual_ref_eq}
r_{ab}(\lda-\nu) \  k_a(\lda)k_b(\nu)  -  k_b(\nu)\  r_{ab}(\lda+\nu)\ k_a(\lda)  =  k_a(\lda)\  r_{ab}(-\lda-\nu) \ k_b(\nu)  -   k_a(\lda) k_b(\nu)\  r_{ab}(-\lda+\nu)\;.
\end{eqnarray}
In the other well-known case where the $r$-matrix depends only on the quotient of the spectral parameters and is skew-symmetric, the choice $g^{(1)}(\nu)=-1$ and $\tau(\nu)=1/\nu$ gives the multiplicative version of the classical reflection equation.
\end{prop}
However, let us emphasize that the classical 2-reflection equation introduced here generalizes the usual ones because of the introduction of the functions $g^{(1)}$ and $\tau$. 
We will see on the examples below that we can get new interesting solutions from this generalized form. One remarkable feature is the possibility to act with M\"obius transformations on the spectral parameter using $\tau$.
\begin{prop} \label{pro:rb} Let $r$ be a solution of the classical Yang-Baxter equation \eqref{eq:cybe} and $k(\nu)$ a solution of the classical $N$-reflection equation \eqref{eq:NRE}. Let us define 
	\begin{equation}\label{eq:rb}
	\overline r_{ab}(\lda,\nu)=\sum_{j=0}^{N-1}\  g^{(j)}(\nu)\  k^{(j)}_b(\nu)\  r_{ab}\left( \lda,\tau^j(\nu)\right) \left( k^{(j)}_b(\nu)\right)^{-1}\;.
	\end{equation}
Then $\overline r$ satisfies the classical Yang-Baxter equation \eqref{eq:cybe}.
\end{prop}
\proof For convenience, for a given $r$-matrix, let us introduce the notation 
\begin{equation}
CYBE(r)_{abc}(\lda,\mu,\nu)= [\  r_{ab}(\lda,\mu)\ ,\  r_{ac}(\lda,\nu)\ ]+[\  r_{ab}(\lda,\mu)\ ,\ r_{bc}(\mu,\nu)\ ] -[\  r_{ac}(\lda,\nu)\ ,\  r_{cb}(\nu,\mu)\ ]\,.
\end{equation}
Note that with our definition of $\overline r$, the classical $N$-reflection equation \eqref{eq:NRE} may be rewritten compactly as
\begin{equation}\label{eq:reb}
\overline r_{ab}(\lda,\nu) k_a(\lda) = k_a(\lda) \overline r_{ab}(\tau(\lda),\nu)\;.
\end{equation}
In turn, replacing $\lda$ by $\tau(\lda)$ in \eqref{eq:reb}, we see that \eqref{eq:reb} implies, 
\begin{equation}\label{eq:unj}
\overline r_{ab}(\lda,\nu) k^{(n)}_a(\lda) = k^{(n)}_a(\lda) \overline r_{ab}(\tau^n(\lda),\nu)\,,~~n=1,2,\dots\;.
\end{equation}
Now, using \eqref{eq:unj} in the form 
\begin{equation}
(k_b^{(n)}(\mu))^{-1}\,\overline r_{bc}(\mu,\nu)=\overline r_{bc}(\tau^n(\mu),\nu)\,(k_b^{(n)}(\mu))^{-1}
\end{equation}
and its counterpart under the exchange $b\leftrightarrow c$, $\mu \leftrightarrow \nu$, we can write
\begin{eqnarray}
&&CYBE(\overline{r})_{abc}(\lda,\mu,\nu)\nonumber\\
&&=\sum_{n,m=0}^{N-1}g^{(n)}(\mu)g^{(m)}(\nu)\  k^{(n)}_b(\mu)k^{(m)}_c(\nu)\,CYBE(r)_{abc}(\lda,\tau^n(\mu),\tau^m(\nu))\,(k^{(m)}_c(\nu))^{-1}(k^{(n)}_b(\mu))^{-1}\;.
\end{eqnarray}
The proposition follows from the fact that $CYBE(r)_{abc}(\lda,\mu,\nu)=0$.
\endproof 
This proposition allows us to construct a new solution $\overline{r}$ of the classical reflection equation for each solution $k$ of the $N$-reflection equation and each solution $r$ of the classical Yang-Baxter equation.
It generalizes the results of \cite{skrypnyk}, stated for the usual reflection equation. In general, the r-matrices $\overline r$ obtained by this construction are not skew-symmetric even if the starting $r$-matrix $r$ is. However, they still allow for the construction of interesting algebraic structures and integrable models as shown in Section \ref{sec:alg}.

\begin{rem}
In the course of the proof, we used a compact form of the $N$-reflection equation which is worth pointing out separately
	\begin{equation}
	\overline r_{ab}(\lda,\nu) k_a(\lda) = k_a(\lda) \overline r_{ab}(\tau(\lda),\nu)\;.
	\end{equation}
\end{rem}

It is convenient to introduce the notion of N-unitary relation which generalizes the so-called unitary relation ($k(\lda)k(-\lda)\propto\1_n$ or $k(\lda)k(1/\lda)\propto\1_n$) often required for the usual reflection equation. 
\begin{defi}
The matrix $k$ is said to satisfy the $N$-unitary relation if
\begin{equation}\label{eq:nuni}
k^{(N)}(\nu)=f(\nu) \1_n \quad\text{i.e.}\qquad k(\nu)\; k(\tau(\nu))\; k(\tau(\tau(\nu)))\dots k(\tau^{N-1}(\nu))=f(\nu) \1_n\;,
\end{equation}
where $f(\nu)$ is a scalar function.
\end{defi}

\begin{rem}
If $\tau^N(\nu)=\nu$, then equation \eqref{eq:unj} for $j=N$ becomes $\overline r_{ab}(\lda,\nu) k^{(N)}_a(\lda) = k^{(N)}_a(\lda) \overline r_{ab}(\lda,\nu)$. This equation is automatically satisfied if the $k$-matrix satisfies the N-unitary relation.
\end{rem}

\begin{rem}\label{rem:sym} If $\tau^N(\nu)=\nu$ and $g^{(j)}(\nu)=\omega^j$ with $\omega^N=1$, then a solution $k$ satisfying the N-unitary relation \eqref{eq:nuni} and the following symmetry relation
\begin{equation}\label{eq:sym}
  r_{ab}(\lda,\nu) = \omega \,k_a(\lda)\,k_b(\nu)\, r_{ab}(\tau(\lda),\tau(\nu))\,k_b(\nu)^{-1}\,k_a(\lda)^{-1}
\end{equation}
is a solution of the $N$-reflection equation. This is a case of a reduction by a $\ZZ_N$ action on the $r$-matrix.
 Equation \eqref{eq:sym} underlies the results obtained in \cite{CY} and is a particular case of the $N$-reflection equation.
\end{rem}

\section{Some solutions of the $N$-reflection equation \label{sec:solu}}

In this section, we present some solutions of the $N$-reflection equation for various values of $N$ and for the standard rational and trigonometric $r$-matrices. Even in the case $N=2$, some of these solutions are new compared to the usual reflection equation and, in fact, cannot be accommodated by it. As explained previously, all these solutions allow us to obtain new solutions 
of the non-standard classical Yang-Baxter equation.

\subsection{Rational r-matrix}\label{rat}

In this subsection, we provide some solutions of the $N$-reflection equation associated to the rational r-matrix 
\begin{equation}\label{eq:rr}
 r(\lda,\mu)=r(\lda-\mu)=\frac{P}{\lda-\mu}\;,
\end{equation}
where P is the permutation operator of $\CC^n \otimes \CC^n$. In this case, the following matrices are solutions of the usual reflection equation \eqref{eq:RE}
\begin{equation}
 k(\lda)=\theta \1_n +\lda G\;,
\end{equation}
where $\theta$ is a free parameter and $G$ is a $n\times n$-matrix satisfying $G^2=\1_n$. This result generalizes to the case of the $N$-reflection equation as follows:
\begin{prop}\label{pro:G}
 The matrix
 \begin{equation}\label{eq:sol1}
 k(\lda)=\theta \1_n +\lda G\;,
\end{equation}
where G is an $n\times n$-matrix satisfying $G^N=\1_n$, is a solution of the $N$-reflection equation with $g^{(j)}(\nu)=\omega^j$, $\tau^j(\nu)=\omega^j \nu$ and $\omega=\exp(2i\pi/N)$.
This solution also satisfies the N-unitary relation \eqref{eq:nuni} with $f(\lda)=\theta^N+\lda^N\exp(i \pi(N-1))$.
\end{prop}
\proof We start by proving the $N$-unitary relation:
\begin{eqnarray}
 k(\lda)k(\omega\lda)\dots k(\omega^{N-1}\lda)=\prod_{j=0}^{N-1} (\theta + \omega^j \lda G)=\theta^N +\lda^N G^N \prod_{j=0}^{N-1} \omega^j\;.
\end{eqnarray}
We have used well-known relations for the roots of unity, such as $\displaystyle \sum_{j=0}^{N-1} \omega^j=0$ or  
$\displaystyle  \sum_{0\leq j<k\leq N-1} \omega^{j+k}=0$. The result follows from  $G^N=\1_n$.

Next, multiplying the $N$-reflection equation by $k_b^{(N-1)}(\nu)$ on the right and using the property of the permutation operator, the $N$-reflection equation is equivalent to
\begin{eqnarray}
 &&\sum_{p=0}^{N-1} \frac{ \omega^p P_{ab}}{\lda-\omega^p \nu} k_a(\nu)\dots k_a(\omega^{p-1}\nu)\ k_a(\lda)\ k_b(\omega^p\nu)\dots k_b(\omega^{N-2}\nu)\nonumber\\
&=& \sum_{p=0}^{N-1}\frac{ \omega^p P_{ab}}{\omega \lda-\omega^p \nu} k_a(\nu)\dots k_a(\omega^{p-1}\nu)\ k_b(\lda)\ k_b(\omega^p\nu)\dots k_b(\omega^{N-2}\nu)\;.
\end{eqnarray}
By multiplying on the left by $P_{ab}$ and rearranging the sum, one gets equivalently
\begin{eqnarray}
 &&\sum_{p=0}^{N-2} \frac{ \omega^p }{\lda-\omega^p \nu} k_a(\nu)\dots k_a(\omega^{p-1}\nu)\ \Big[k_a(\lda)k_b(\omega^p\nu)- k_a(\nu \omega^p) k_b(\lda) \Big]\ k_b(\omega^{p+1}\nu)\dots k_b(\omega^{N-2}\nu)\nonumber\\
&=& \frac{ \omega^{N-1} }{\lda-\omega^{N-1} \nu} \Big(  k_b(\lda)\ k_b(\nu)\dots k_b(\omega^{N-2}\nu) -k_a(\nu)\dots k_a(\omega^{N-2}\nu)\ k_a(\lda) \Big)  \;.\label{eq:reeq}
\end{eqnarray}
We remark that the expression inside the square bracket in the L.H.S. of the previous relation can be written as follows by using the explicit form \eqref{eq:sol1} of $k$
\begin{equation}
 k_a(\lda)k_b(\omega^p\nu)- k_a(\nu \omega^p) k_b(\lda)= \frac{\theta(\lda-\omega^p\nu)}{\omega^p\nu}\left( k_a(\omega^p\nu)-k_b(\omega^p\nu) \right)\;.
\end{equation}
Then, we recognize a telescopic sum in the L.H.S. of \eqref{eq:reeq} and the $N$-reflection equation reduces to
\begin{eqnarray}
 &&\frac{\theta}{\nu}( k_a(\nu)\dots k_a(\omega^{N-2}\nu)- k_b(\nu)\dots k_b(\omega^{N-2}\nu))\nonumber\\
&=& \frac{ \omega^{N-1} }{\lda-\omega^{N-1} \nu} \Big(  k_b(\lda)\ k_b(\nu)\dots k_b(\omega^{N-2}\nu) -k_a(\nu)\dots k_a(\omega^{N-2}\nu)\ k_a(\lda) \Big)  \;.\label{eq:reeq2}
\end{eqnarray}
By considering separately the terms in the space $a$ and the ones in the space $b$ and by using the $N$-unitary relation proved previously, we obtain that relation \eqref{eq:reeq2} holds, which proves 
that the $N$-reflection equation is satisfied.
\endproof
For $\theta=0$, the $k$-matrix \eqref{eq:sol1} satisfies the stronger relation \eqref{eq:sym} and has been studied previously (see e.g. \cite{CY}). 

The importance of the functions $\tau$ and $g$ can be seen from the existence of the following solutions of the $N$-reflection equation.
\begin{prop}\label{pro:2s}
 The identity matrix is a solution of the $2$-Reflection equation with
 \begin{equation}\label{eq:re2}
  \tau(\nu)=\frac{a\nu+b}{c\nu-a}\quad \text{and}\qquad g^{(1)}(\nu)=-\frac{a^2+bc}{(a-c\nu)^2}\;,
 \end{equation}
 where $a$, $b$ and $c$ are free parameters such that $\tau\neq 0,\infty$ identically. In particular, $\tau^2(\nu)=\nu$.
 
 The identity matrix is a solution of the $3$-Reflection equation with
 \begin{equation}\label{eq:re3}
  \tau(\nu)=\frac{a\nu+b}{c\nu+d}\quad\text{,}\quad g^{(1)}(\nu)=\frac{ad-bc}{(d+c\nu)^2}\quad\text{and}\qquad g^{(2)}(\nu)=\frac{ad-bc}{(a-c\nu)^2}\;,
 \end{equation}
  where $a$, $b$, $c$ and $d$ are parameters constrained by $a^2+ad+bc+d^2=0$ and such that such that $\tau\neq 0,\infty$ identically. In particular, $\tau^3(\nu)=\nu$.
\end{prop}
\proof The $N$-reflection equation for the rational $r$-matrix and for a $k$-matrix equals to the identity matrix reduces to one functional equation
\begin{equation}
 \sum_{j=0}^{N-1} g^{(j)}(\nu) \frac{1}{\lda-\tau^j(\nu)}=\sum_{j=0}^{N-1} g^{(j)}(\nu) \frac{1}{\tau(\lda)-\tau^j(\nu)}\;.
\end{equation}
One can show by direct computation that for the choices \eqref{eq:re2} and \eqref{eq:re3}, the last equation holds. This finishes the proof. 

\endproof
 Again, this proposition allows for the construction of new solutions of the non-standard classical Yang-Baxter equation where the poles of the $r$-matrix are governed by the parameters 
of the function $\tau$. For $b=c=0$ and in the case $N=2$, we recover the map $\tau(\nu)=-\nu$ usually used for the reflection equation in the rational case. Despite some effort, we have not been able to 
produce a solution where $\tau$ is a M\"obius transformation and $k$ is not the identity matrix in the rational case. However, below we present examples where this is possible in the trigonometric case.

\subsection{Trigonometric r-matrix}

In this subsection, we study the $N$-reflection equation in the case of the $4\times 4$ trigonometric solution of the classical Yang-Baxter equation \eqref{eq:cybe} defined by 
\begin{equation}
 r(\lda,\nu)=r(\lda/\nu)=\frac{1}{2(\lda-\nu)}\begin{pmatrix}
       -\lda-\nu & 0 & 0 & 0\\
       0&\lda+\nu &-4\nu &0\\
       0&-4\lda &\lda+\nu &0\\
       0 & 0 &0 & -\lda-\nu
      \end{pmatrix}\;.
\end{equation}

\begin{prop}
The 2-reflection equation with
 \begin{equation}
  \tau(\nu)=\frac{a\nu+b}{c\nu-a}\quad,\qquad g^{(1)}(\nu)=-\; \frac{(a^2-bc)\nu}{(c\nu-a)(a\nu+b)}
 \end{equation}
 has two solutions: the identity matrix and the following diagonal matrix
 \begin{equation}
  k(\nu)=\begin{pmatrix}
         \tau(\nu)&0\\
          0 & \nu
         \end{pmatrix}\;.
 \end{equation}
The 3-reflection equation with
 \begin{equation}
  \tau(\nu)=\frac{a\nu+b}{c\nu+d}\ , \quad g^{(1)}(\nu)=\; \frac{(a+d)^2\nu}{(a\nu+b)(c\nu+d)}\quad,\qquad 
  g^{(2)}(\nu)=\; \frac{(a+d)^2\nu}{(d\nu-b)(-c\nu+a)}
 \end{equation}
and the constraint $a^2+bc+ad+d^2=0$ has five solutions: the identity matrix and the following diagonal matrices
 \begin{eqnarray}
 && k(\nu)=\begin{pmatrix}
          \tau(\nu)&0\\
          0 & \nu
         \end{pmatrix}\ , \quad         
   k(\nu)=\begin{pmatrix}
          \tau^2(\nu)&0\\
          0 & \nu
         \end{pmatrix}\ ,\quad k(\nu)=\begin{pmatrix}
         \tau(\nu)&0\\
          0 & \tau^2(\nu)
         \end{pmatrix}\;,\\      
   &&k(\nu)=\begin{pmatrix}
          (a+d)(a\nu+b)&0\\
          0 & (c\nu-a)(d\nu-b)
         \end{pmatrix}\ ,\quad k(\nu)=\begin{pmatrix}
          (c\nu-a)(d\nu-b)&0\\
          0 & (a+d)(c\nu+d)
         \end{pmatrix}.\qquad
 \end{eqnarray}
\end{prop}
\proof The proposition is proven by direct computation.
\endproof
In the particular case $a=0$ and $b=c$, the 2-reflection equation becomes the usual one in the case of the trigonometric $r$-matrix since $\tau(\nu)=1/\nu$. 

\section{Integrable models based on the $N$-reflection equation \label{sec:alg}}

\subsection{Poisson subalgebra and Lax pair of the equations of motion}

Let us consider the following Poisson algebra,
\begin{equation}
\label{rll}
 \{L_a(\lambda)\ ,\ L_b(\mu)\}=\left[ r_{ab}(\lambda,\mu)\ , \ L_a(\lambda) + L_b(\mu) \right]\,,
\end{equation}
associated to a skew-symmetric matrix $r$ solution of the classical Yang-Baxter equation and define
\begin{equation}\label{eq:B}
 B_a(\lda)= \sum_{j=0}^{N-1} g^{(j)}(\lda) k^{(j)}_a(\lda)  L_a(\tau^j(\lda))   \left( k^{(j)}_a(\lda)\right)^{-1}\;.
\end{equation}
\begin{prop}
	The elements $B(\lda)$ form a Poisson subalgebra of the form
\begin{equation}
\label{rbb}
\{B_a(\lambda)\ ,\ B_b(\mu)\}= \left[ \overline r_{ab}(\lambda,\mu)\ , \ B_a(\lda) \right] - \left[ \overline r_{ba}(\mu,\lambda)\ , \ B_b(\nu) \right]\,,
\end{equation}
where $\overline r$ is given by \eqref{eq:rb} and $k$ is a solution of the $N$-reflection equation.
\end{prop}
\proof 
This is shown by direct computation starting from \eqref{rll} and using the definition of $B(\lda)$ \eqref{eq:B}. If one uses
the $N$-reflection equation, one can bring all the terms produced in the right-hand side into the simple form given in \eqref{rbb}.  The skew-symmetry of the Poisson bracket \eqref{rbb} is evident and 
the Jacobi identity is a consequence of the classical Yang-Baxter equation \eqref{eq:cybe} satisfied by $\overline r$.  The latter fact is another motivation for 
introducing the $N$-reflection equation in the first place.
\endproof
The Poisson algebra \eqref{rbb} provides an example of a linear Poisson structure based on a non skew-symmetry $r$-matrix, as discussed in \cite{BV} (see also \cite{Maillet}). 

\begin{rem}
 If the hypotheses of Remark \ref{rem:sym} hold, then 
 \begin{equation}
  \phi \ :\ L(\lda)\mapsto \omega\ k(\lda) L(\tau(\lda)) k(\lda)^{-1}
 \end{equation}
is a $\ZZ_N$ action on the Poisson algebra \eqref{rll} and $B_a(\lda)$ contains the generators of the fixed point subalgebra of this action. 
\end{rem}

\begin{rem}
Relation \eqref{eq:B} looks similar to the reduction group scheme for the Lax operator considered in \cite{Mik,Mik2,MOP,RF,AT,Av1} in which $k$ and $\tau$ provide a representation of the reduction group.
In our case, the approach is different in that we do not assume {\it a priori} that $k$ and $\tau$ are associated to a group. Instead, we look for conditions on them ensuring that the Poisson algebra for $B(\lda)$ closes. This leads to our N-reflection equation \eqref{eq:NRE}. It is an open and interesting problem to study the interplay between the N-reflection equation and the reduction group approach.
\end{rem}

It is well-known \cite{BV} that the Poisson algebra \eqref{rbb} allows for the construction of Poisson commuting elements since, for $p,q=1,2,\dots$, one gets
\begin{equation}\label{eq:comm}
 \{ tr B(\lambda)^p\ , \ tr B(\nu)^q\}=0\;.
\end{equation}
If we choose as Hamiltonian\footnote{Here our terminology is rather general and by ``Hamiltonian'', we mean any element that lives 
	in the abelian subalgebra generated by $tr B(\lambda)^p$.}  $tr B(\lambda)^p$, then the equations of motion have a Lax representation
\begin{equation}
 \{ tr B(\lambda)^p,B(\nu)\}= \dot  B(\nu) =  [ B(\nu), M(\lambda,\nu)]
\end{equation}
with 
\begin{equation}
 M_b(\lda,\nu)=p \ tr_a\left( B_a(\lambda)^{p-1} \overline r_{ba}(\nu,\lda) \right)\;.
\end{equation}
The matrix $M(\lda,\nu)$ satisfies the important relation
\begin{equation}
\label{KMMK}
 M(\lda,\nu) k(\nu)=k(\nu) M(\lda,\tau(\nu))\;.
\end{equation}
This relation is a straightforward consequence of the $N$-reflection equation written as in \eqref{eq:reb}.
It is a generalization of the relation introduced in the context of the usual classical reflection equation \cite{Sk} (which is recovered as explained after \eqref{eq:NRE2}). 
We recall that in that standard context, relation \eqref{KMMK} can be taken as a definition of integrable boundary conditions. In \cite{ACC}, the Hamiltonian interpretation of this relation (and its generalisation to the dynamical case) was presented in the case of the quadratic Poisson bracket and the usual reflection equation. It is an open problem to generalise the results of \cite{ACC} to the present case of the $N$-reflection equation.

\subsection{Gaudin models}

In this section, we apply the previous results to the construction of new integrable models which are generalizations of the Gaudin models \cite{Gaudin}. The general procedure goes as follows.
Suppose we are given a local representation of \eqref{rll} in the form\footnote{We restrict ourselves here to the case of additive spectral parameters.}, for $m,p=1,2,\dots,L$,
\begin{equation}\label{rll_local}
 \{\ell_a(m,\lambda)\ ,\ \ell_b(p,\mu)\}=\delta_{mp}\left[ r_{ab}(\lambda-\mu)\ , \ \ell_a(m,\lambda) + \ell_b(p,\mu) \right]\,.
\end{equation}
Given a solution $k$ of the $N$-reflection equation, we obtain a representation of relation \eqref{rbb} by setting
\begin{equation}
 B_a(\lda)= \sum_{m=1}^L   \sum_{j=0}^{N-1} g^{(j)}(\lda) k^{(j)}_a(\lda) \ell_a(m,\tau^j(\lda)-z_m)    \left( k^{(j)}_a(\lda)\right)^{-1}\;,
\end{equation}
where $z_m$ are free, mutually distinct, parameters.
The Gaudin-type Hamiltonians are then defined by, 
\begin{equation}\label{eq:Hg}
 H_m=\frac{1}{2}\text{res}_{\lda=z_m}\ tr_a \left(B_a(\lda)\right)^2\,,~~m=1,2,\dots,L \;.
\end{equation}
These Hamiltonians are in involution due to relation \eqref{eq:comm} and allow us to define an integrable model.
Therefore, each solution of the $N$-reflection equation allows us to get integrable Gaudin-type models based on a non-skew $r$-matrix $\bar{r}$. 
In that sense, our construction produces explicit classes of models falling into the general scheme of \cite{Skryp} and generalizes the method used in \cite{skrypnyk}.

\subsection{Explicit example}
We now present an explicit example associated to the 2-reflection equation in the rational case for $n=2$. 
Let us introduce $L$ copies of the $su(2)$ Poisson algebra with generators $\{s_j^+, s_j^- , s_j^z\}$ satisfying
\begin{equation}
 \{ s_j^+\ , s_k^-\}=  \delta_{jk} s_j^z\quad,\qquad  \{s_j^z\ , s_k^\pm\}=\pm 2 \delta_{jk} s_j^\pm \quad\text{and}\qquad \frac{1}{2}(s^z_j)^2 +2s^+_j s^-_j= \mathfrak{s}_j^2\;,
\end{equation}
where $\mathfrak{s}_j$ are some parameters. 
Then 
\begin{equation}
 \ell(j,\nu)=\frac{1}{\nu}\begin{pmatrix} 
             \frac{1}{2} s^z_j & s^+_j\\
              s^-_j & -\frac{1}{2}s^z_j
             \end{pmatrix}\;,
\end{equation}
satisfies \eqref{rll_local} with the rational r-matrix \eqref{eq:rr}. 
From the solution of the $2$-reflection equation given in Proposition \ref{pro:2s}, Proposition \ref{pro:rb} provides the following r-matrix
\begin{equation}\label{eq:rro2}
 \overline r(\lda,\mu)=P\left(\frac{1}{\lda-\mu}-\frac{(a^2+bc)}{(a-c\mu)(b+a(\lda+\mu)  -c\lda\mu) }\right)\;.
\end{equation}
Note that the r-matrix \eqref{eq:rro2} is related to the rational r-matrix \eqref{eq:rr} via the transformation
 \begin{equation}
  \overline r(\lda,\mu)=\frac{b+2a\mu-c\mu^2}{2(a-c\mu)^2}\ r( p(\lda)-p(\mu) )
 \end{equation}
where $p(\mu)= \frac{b+c\mu^2}{2c(a-c\mu)}$. This is an example of the equivalence relations discussed for instance in \cite{Skrypnyk}.
The Gaudin-type Hamiltonians \eqref{eq:Hg} associated to this r-matrix read
\begin{equation}
H_i=\sum_{\genfrac{}{}{0pt}{1}{k=1}{k\neq i} }^L S_{ik}
\left( \frac{1}{z_i-z_k} +\frac{a^2+bc}{(a-cz_i)( b+a(z_i+z_k)-cz_iz_k)} \right) 
+\frac{(a^2+bc)\mathfrak{s}_i^2}{(a-cz_i)(b+2az_i-cz_i^2)}\;,
\end{equation}
where $S_{ik}=\left(\frac{1}{2}s^z_is^z_k +s^+_i s^-_k +s^-_i s^+_k\right) $.
Note that in the case $b=c=0$, we find the Gaudin Hamiltonians based on the $BC_L$ root system \cite{Hik}, consistently with the remark at the end of Section \ref{rat}.

Similarly, from the solution of the $3$-reflection equation given in Proposition \ref{pro:2s}, we obtain the following r-matrix with Proposition \ref{pro:rb}
(we recall that in this case $a^2+bc+ad+d^2=0$)
\begin{equation}\label{eq:rro3}
 \overline r(\lda,\mu)=P\left(\frac{1}{\lda-\mu}-\frac{ad-bc}{(d+c\mu)( b+a\mu-d\lda-c\mu\lda)}+\frac{ad-bc}{(a-c\mu)( b+a\lda-d\mu-c\mu\lda)} \right) \;.
\end{equation}
This r-matrix \eqref{eq:rro3} can be also obtained from the following transformation of the rational r-matrix  \eqref{eq:rr}
\begin{equation}
  \overline r(\lda,\mu)=\frac{b^2(c\mu-a)^2(c\mu+d)^2}{(c\mu^2-(a-d)\mu-b)^2}\  r( p(\lda)-p(\mu) )
 \end{equation}
where $p(\mu)= \frac{c^3\mu^3-(a-d)^2c^2\mu^2-c(a+2d)(2a+d)\mu+(a-d)(a+d)^2}{c^3b^2(c\mu-a)(c\mu+d)}$.
In this case, the Gaudin-type Hamiltonians \eqref{eq:Hg} read
\begin{eqnarray}
H_i&=&\sum_{\genfrac{}{}{0pt}{1}{k=1}{k\neq i} }^L S_{ik}
\left( \frac{1}{z_i-z_k} +\frac{ad-bc}{(d+cz_i)( b+az_i-dz_k-cz_iz_k)}-\frac{ad-bc}{(a-cz_i)( b+az_k-dz_i-cz_iz_k)} \right) \nonumber\\
&&+\frac{(ad-bc)\mathfrak{s}_i^2}{b+(a-d)z_i-cz_i^2)}\left(\frac{1}{d+cz_i}-\frac{1}{a-cz_i}\right)\;,
\end{eqnarray}
In the case $b=c=0$ and $\frac{a}{d}=\omega=e^{\frac{2i\pi}{3}}$, this reduces to a $\ZZ_3$-cyclotomic classical Gaudin Hamiltonians
\begin{eqnarray}
H_i&=&\sum_{\genfrac{}{}{0pt}{1}{k=1}{k\neq i} }^L S_{ik}
\left( \frac{1}{z_i-z_k}+\frac{1}{z_i-\omega z_k}+\frac{1}{z_i-\omega^2 z_k}  \right) +\frac{\mathfrak{s}_i^2}{z_i}\;,
\end{eqnarray}
whose quantum counterparts were introduced in \cite{CY} and further developed in \cite{VY}.

\section{Conclusions and outlook}

We introduced the $N$-reflection equation as a generalization of the usual reflection equation \cite{Sk}. An important motivation
is the possibility to define a consistent Poisson algebra \eqref{rbb} when one applies the following map  
\begin{equation}
\label{mapN}
L_a(\lda)\mapsto \sum_{j=0}^{N-1} g^{(j)}(\lda) k^{(j)}_a(\lda)  L_a(\tau^j(\lda))   \left( k^{(j)}_a(\lda)\right)^{-1}
\end{equation}
on the Poisson algebra 
\begin{equation}
\{L_a(\lambda)\ ,\ L_b(\mu)\}=\left[ r_{ab}(\lambda,\mu)\ , \ L_a(\lambda) + L_b(\mu) \right]\,.
\end{equation}
The map \eqref{mapN} appears as a generalization of the maps
\begin{equation}
L_a(\lda)\mapsto L_a(\lda)\pm k_a(\lda)  L_a(\sigma(\lda))k_a(\lda)^{-1}
\end{equation}
producing the eigenspaces of the involution
\begin{equation}
\label{involution}
L_a(\lda)\mapsto k_a(\lda)  L_a(\sigma(\lda))k_a(\lda)^{-1}
\end{equation}
where $k(\lda)\,k(\sigma(\lda))=\1$ and $\sigma(\lda)=-\lda$ or $\sigma(\lda)=1/\lda$. These are known to be related to the usual classical reflection equation \eqref{usual_ref_eq} (or its multiplicative form). But in our 
case, we do not necessarily require a priori that there is an underlying map of order $N$ which would be the generalization of the involution \eqref{involution}.
Several consequences of this construction are noteworthy. Firstly, it gives a systematic way to construct many examples of non skew-symmetry 
$r$-matrices. In this respect, we note that 
once such an $r$-matrix has been obtained, it can be taken as the starting point to repeat the procedure. It is an interesting question whether this process terminates or not and under which conditions on $g^{(j)}$, $\tau$ and $k$. The relationship between all the solutions of the classical Yang-Baxter equation that can be obtained in this way and the entire set of such solutions is also an open problem which would require an understanding of the full classification of non skew-symmetric solutions of that equation. Such a classification is not known so far.  Secondly, one can define new Gaudin models in a systematic way. Despite the possibility to relate the $r$-matrix of the new models to the standard rational $r$-matrix via rescaling and reparametrization, the nature of these models is quite different from the standard Gaudin models. Indeed, rescaling and reparametrization of an $r$-matrix are known to have consequences on the model, one of which is that reparametrization affect the skew-symmetry of the $r$-matrix, hence the (non) ultralocality of the model (in field theoretical language). In our case, these operations depends on the M\"obius transformations acting on the spectral parameters and it is the first time that such transformations are used in this context to our knowledge.
Thirdly, our results give a natural context for \eqref{KMMK} in cases where $\tau$ is not of order $2$. Such examples have appeared before in the context of linearizable boundary conditions \cite{Fokas}. However, there is no hint of an underlying Hamiltonian description of \eqref{KMMK} in that context. We have provided such a Hamiltonian formulation in what we could call a time-independent setting. We hope that
our results can be incorporated into a larger theory which would contain \eqref{KMMK} as a special case, along the same lines as the results obtained in  \cite{ACC} in the standard reflection equation case. A major step in achieving this would be to find the appropriate dynamical 
generalization of our $N$-reflection equation. This is a completely open problem. 

A related open question is the problem of quantization of the $N$-reflection equation in the spirit of the quantization of the classical reflection equation \cite{Skquant,FrMa}. In that context, it is known that the semi-classical limit of the quantum structure gives the appropriate Poisson algebra on the matrices $k$.

\paragraph{Acknowledgment:} N. Cramp\'e acknowledges the hospitality 
of the School of Mathematics, University of Leeds where this work was completed. N. Cramp\'e's visit was partially supported by the Research Visitor Centre of the School of Mathematics. 
We are grateful to J.~Avan, A.~Molev, E.~Ragoucy and T.~Skrypnyk for their comments.


\begin{thebibliography}{99}

\bibitem{Skly} E.K. Sklyanin, \textit{Method of the inverse scattering problem
	and quantum nonlinear Schr\"odinger equation}, Dokl. Acad. Nauk SSSR
{\bf 244} (1978), 1337.

\bibitem{STS1} M.A. Semenov-Tian-Shansky, \textit{What is a classical r-matrix?}, Funct. Anal. Appl. {\bf 17} (1983), 259.

\bibitem{FT}
L.D. Faddeev, L.A. Takhtadjan, {\it Hamiltonian Methods in the Theory of Solitons},
Springer-Verlag 1987.

\bibitem{Sk} E.K. Sklyanin, 
{\it Boundary conditions for integrable equations,}
Funktsional.  Anal.  i  Prilozhen. \textbf{21} (1987), 86
(English transl.: Funct. Anal. Appl. \textbf{21} (1987), 164). 

\bibitem{CC2} V. Caudrelier, N. Crampe, {\it Integrable N-particle Hamiltonians with Yangian or Reflection Algebra Symmetry}, J.Phys. {\bf A37} (2004), 6285.

\bibitem{CZ1} V. Caudrelier, Q.C. Zhang, {\it Vector Nonlinear Schr\"odinger Equation on the half-line}, J. Phys. {\bf A45} (2012), 105201.

\bibitem{CZ2} V. Caudrelier, Q.C. Zhang, {\it Yang-Baxter and reflection maps from vector solitons with a boundary},  Nonlinearity {\bf 27} (2014), 1081.

\bibitem{poly}  A.P. Polychronakos,
{\it Generalized Calogero models through reductions by discrete symmetries,}
Nucl. Phys. \textbf{B 543} (1999) 485.

\bibitem{DO} C.F. Dunkl, E.M. Opdam,
 {\it Dunkl operators for complex reflection groups,}
Proc. London Math. Soc. \textbf{86} (2003) 70.

\bibitem{CY} N. Crampe, C.A.S. Young,
{\it Integrable Models From Twisted Half Loop Algebras,}
 J. Phys. \textbf{A40} (2007) 5491. 

 \bibitem{CY2} N. Crampe, C.A.S. Young,
 {\it Sutherland models for complex reflection groups,}
Nucl. Phys. \textbf{B 797} (2008) 499.

\bibitem{CC} V. Caudrelier, N. Crampe, {\it Symmetries of Spin Calogero Models},
 SIGMA {\bf 4} (2008), 090.

\bibitem{EF}P. Etingof, G.Felder, X.Ma, A. Veselov,
{\it On elliptic Calogero--Moser systems for complex crystallographic reflection groups,}
J. Algebra \textbf{329} (2011) 107.

\bibitem{VY} B. Vicedo, C.A.S. Young,
{\it Cyclotomic Gaudin models: construction and Bethe ansatz,} 
 Commun. Math. Phys. \textbf{343} (2016) 971. 


\bibitem{CS} O. Chalykh, A. Silantyev,
{\it KP hierarchy for the cyclic quiver},
J. Math. Phys. \textbf{58} (2017) 071702.

\bibitem{CF}O. Chalykh, M. Fairon,
{\it  Multiplicative quiver varieties and generalised Ruijsenaars-Schneider models,}
 \texttt{arXiv:1704.05814}.
   
\bibitem{skrypnyk} T. Skrypnyk,
{\it Generalized Gaudin spin chains, nonskew symmetric r-matrices, and reflection equation algebras,}
J. Math. Phys. \textbf{48} (2007) 113521.     

\bibitem{BV} O. Babelon, C.M. Viallet,
{\it Hamiltonian structures and Lax equations,} Phys.Lett \textbf{B237} (1989) 411. 

\bibitem{Maillet} J-M. Maillet, {\it New integrable canonical structures in two-dimensional models}, Nucl. Phys. {\bf B269} (1986) 54.

\bibitem{Mik} A.V. Mikhailov, {\it Integrability of a two-dimensionnal  generalization of the Toda chain,} JETP Lett. \textbf{30} (1979) 414.

\bibitem{Mik2} A.V. Mikhailov, {\it The reduction problem and the inverse scattering method,} Physica \textbf{3D} (1981) 73.

\bibitem{MOP} A.V. Mikhailov, M.A. Olshanetsky, A.M. Perelomov, {\it Two-dimensional generalized Toda lattice,} Commun. Math. Phys. \textbf{79} (1981) 473.

\bibitem{RF}
N.Y. Reshetikhin, L.D. Faddeev, {\it Hamiltonian structures for integrable models of field theory,} Theor. Math. Phys. \textbf{56} (1983) 847.

\bibitem{AT} J. Avan, M. Talon, {\it Graded $R$-matrices for integrable systems,} Nucl. Phys. \textbf{B 352} (1991) 215.

\bibitem{Av1} J. Avan, {\it From rational to trigonometric R-matrices,} Phys. Lett. \textbf{ A 156} (1991) 61.



\bibitem{ACC} J. Avan, V. Caudrelier, N. Crampe, {\it From Hamiltonian to zero curvature formulation for classical integrable boundary conditions,}
 \texttt{ arXiv:1802.07593}.   

 \bibitem{Gaudin} M. Gaudin,
{\it Diagonalisation d'une classe d'hamiltoniens de spin,}
J. Phys. France \textbf{37} (1976) 1087. 

\bibitem{Skryp} T. Skrypnyk, {\it New integrable Gaudin-type systems, classical r-matrices
	and quasigraded Lie algebras}, Phys. Lett. {\bf A334} (2005) 390.
	 
 
 \bibitem{Skrypnyk} T. Skrypnyk, {\it Infinite-dimensional Lie algebras, classical r-matrices, and Lax operators: Two approaches}, J; Math. Phys. {\bf 54} (2013) 103507.
 
\bibitem{Hik} K. Hikami, {\it Separation of variables in the BC-type Gaudin magnet}, J. Phys. {\bf A28} (1995) 4053.

\bibitem{Fokas} A.S. Fokas, {\it Integrable Nonlinear Evolution Equations on the Half-Line}, Comm. Math. Phys. {\bf 230} (2002) 1.

\bibitem{Skquant} E.K. Sklyanin, 
{\it Boundary conditions for integrable quantum systems,}
J. Phys. \textbf{A21} (1988), 2375. 

\bibitem{FrMa} L. Freidel, J.M. Maillet, {\it Quadratic algebras and integrable systems,} Phys. Lett.\textbf{B26} (1991) 278.






\end{thebibliography}
\end{document}